\documentclass[preprint, prx]{revtex4}
\usepackage{amsmath}
\usepackage{amsfonts}
\usepackage{amssymb}
\usepackage{color}
\usepackage{graphicx,float}

\begin{document}

\title{Pattern formation by curvature-inducing proteins on spherical membranes}
\author{Jaime Agudo-Canalejo}
\affiliation{Theory \& Bio-Systems Department, Max Planck Institute of Colloids and Interfaces, 14424 Potsdam, Germany}
\affiliation{Rudolf Peierls Centre for Theoretical Physics, University of Oxford, Oxford OX1 3NP, United Kingdom}
\affiliation{Department of Chemistry, The Pennsylvania State University, University Park, Pennsylvania 16802, United States}
\author{Ramin Golestanian}
\affiliation{Rudolf Peierls Centre for Theoretical Physics, University of Oxford, Oxford OX1 3NP, United Kingdom}
\affiliation{Max Planck Institute for the Physics of Complex Systems, N\"othnitzer Str. 38, D-01187 Dresden, Germany}

\date{\today}

\begin{abstract}
Spatial organisation is a hallmark of all living cells, and recreating it in model systems is a necessary step in the creation of synthetic cells. It is therefore of both fundamental and practical interest to better understand the basic mechanisms underlying spatial organisation in cells. In this work, we use a continuum model of membrane and protein dynamics to study the behaviour of curvature-inducing proteins on membranes of spherical shape, such as living cells or lipid vesicles. We show that the interplay between curvature energy, entropic forces, and the geometric constraints on the membrane can result in the formation of patterns of highly-curved/protein-rich and weakly-curved/protein-poor domains on the membrane. The spontaneous formation of such patterns can be triggered either by an increase in the average density of curvature-inducing proteins, or by a relaxation of the geometric constraints on the membrane imposed by the membrane tension or by the tethering of the membrane to a rigid cell wall or cortex. These parameters can also be tuned to select the size and number of the protein-rich domains that arise upon pattern formation. The very general mechanism presented here could be related to protein self-organisation in many biological processes, ranging from (proto)cell division to the formation of membrane rafts.
\end{abstract}

\maketitle

\section{Introduction}

Spatial organisation into inhomogeneous patterns is an essential feature of living organisms, from the macroscale to the cellular level. In the later case, organisation of the plasma membrane and the cytoplasm into specialised domains is more commonly referred to as cell polarity. \cite{drub96,nels03} This spatial organisation of the cell is necessary in order to coordinate important processes such as cell division, differentiation, or directed cell migration.

As early as in 1952, Turing realised \cite{turi52} that very simple systems that are initially in a spatially homogeneous state can spontaneously self-organise into spatially inhomogeneous patterns. However, it is generally believed \cite{drub96,nels03} that the generation of polarity in cells is the result of a tightly-controlled orchestration involving complex signalling networks and active processes such as the reorganisation of the cellular cytoskeleton. Nevertheless, active systems such as the cytoskeleton have been shown to undergo simple pattern formation, \cite{tham14} and there also exist cells for which polarisation is presumably not generated by the cytoskeleton. \cite{zapu08,leav09,merc14,osaw08,shlo09,schw12,petr15} The underlying mechanisms in these systems are however not well understood.

Very recently, \cite{garc15} a system was identified in which cell polarisation appears to be controlled by a relatively simple pattern-formation mechanism. In the coccal bacterium \emph{Staphylococcus aureus}, essential proteins involved in lipid metabolism were seen to distribute in inhomogeneous spatial patterns, that could be explained by a model that considers the dynamics of curvature-inducing proteins on a spherical membrane. However, the model first introduced in Ref.~\citenum{garc15} is very general, and we expect that it might be able to describe the formation of protein patterns on the surface of other types of cells, as well as in model systems consisting of lipid vesicles and proteins. In this work, we will explore in full generality and detail the predictions of such a model.

\begin{figure}[h]
\centering
\includegraphics[width=0.8\linewidth]{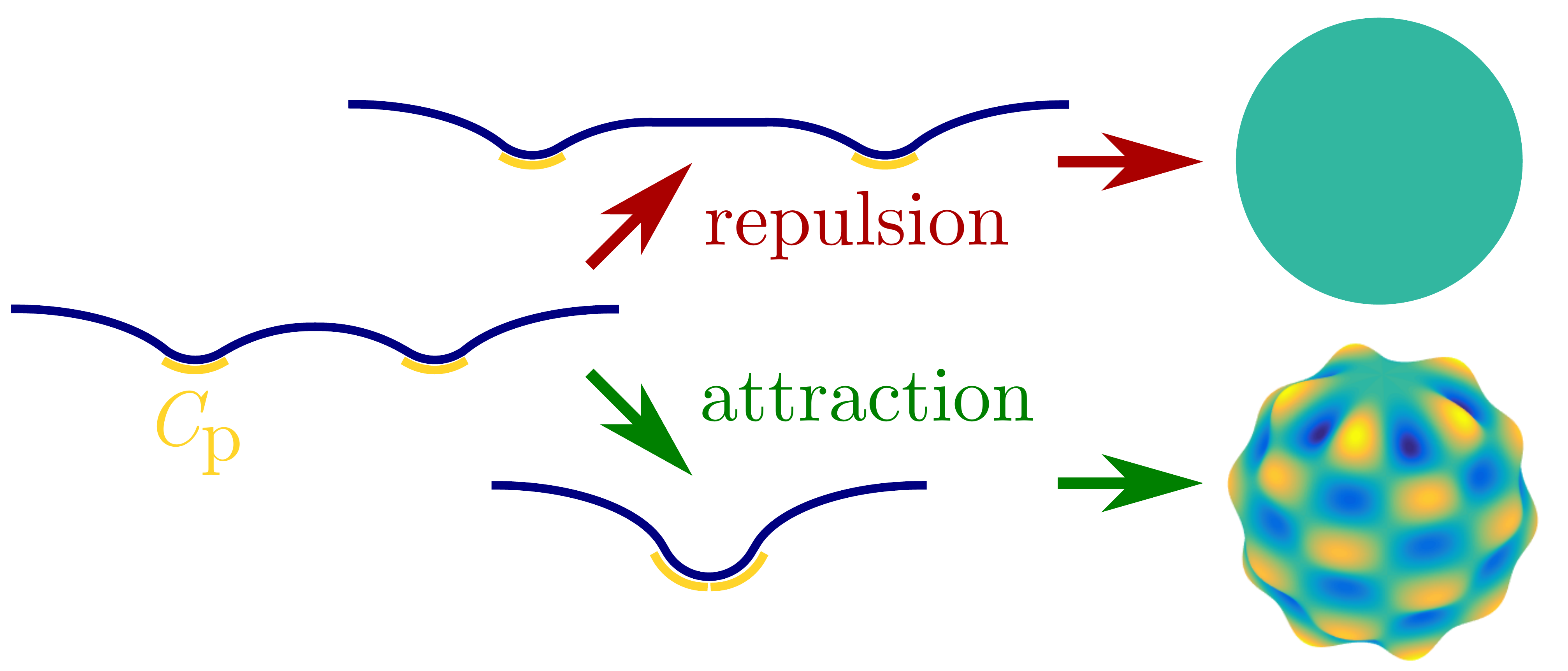}
\caption{The proteins (yellow) impose a spontaneous curvature $C_\mathrm{p}$ on the membrane (blue). Depending on the interplay between curvature, entropic, and membrane tethering/tension forces, proteins might repel, resulting in a spatially homogeneous spherical membrane (turquoise), or they might attract, leading to the spontaneous formation of inhomogeneous patterns of membrane curvature and protein density.}
   \label{fig:scheme}
\end{figure}

The basic idea behind the model is presented in figure~\ref{fig:scheme}. A closed, initially spherical membrane contains proteins that impose a spontaneous curvature $C_\mathrm{p}$ on the membrane (in general, the proteins might be attached to the membrane from the cytoplasmic or the exoplasmic sides, or they might be transmembrane proteins embedded in the membrane). \cite{mcma05,zimm06} If the proteins did not induce any curvature, a random, homogeneous distribution of proteins would be favoured by thermal fluctuations, that is, entropic forces (in the absence of direct attractive protein-protein interactions). However, if the curvature induced by the proteins is large enough, bending contributions to the free energy of the system can lead to an effective attraction between proteins and to the formation of spatially inhomogeneous patterns in protein distribution and membrane curvature. The details of membrane-mediated protein-protein interactions have been thoroughly studied in the past. \cite{gole96a,gole96b,weik98,reyn11} Furthermore, we will consider the possibility of geometric constraints on the membrane, such as the tethering of the membrane to a rigid cell wall/cortex or the existence of a membrane area reservoir at non-zero tension. Interestingly, it was recently shown that solid particles such as proteins can sense the local membrane curvature imposed by geometric constraints on the membrane. \cite{agud17}

Here, we have found that, in realistic situations, spontaneous pattern formation can be induced either by an increase in the surface density of curvature-inducing proteins, or by a decrease in the strength of the geometric constraints on the membrane. Furthermore, these two parameters can also control the size and number of protein-rich (highly curved) and protein-poor (weakly curved) domains. These mechanisms could be exploited by cells in order to trigger spatial organisation of the plasma membrane on demand, and could in principle be replicated in artificial model systems.

The paper is organised as follows. In section~\ref{sec:methods}, we present the continuum model for the energetics and dynamics of the system, and examine the linear stability of the dynamical equations for the shape of the membrane and the protein density distribution. In section~\ref{sec:results}, we explore spontaneous pattern formation in the system as a function of all relevant parameters. Finally, in section~\ref{sec:discussion} we discuss the applicability and consequences of our results in real biological or biomimetic systems.

\section{Methods \label{sec:methods}}

\subsection{Energetics}

\begin{figure}[h]
\centering
\includegraphics[width=0.4\linewidth]{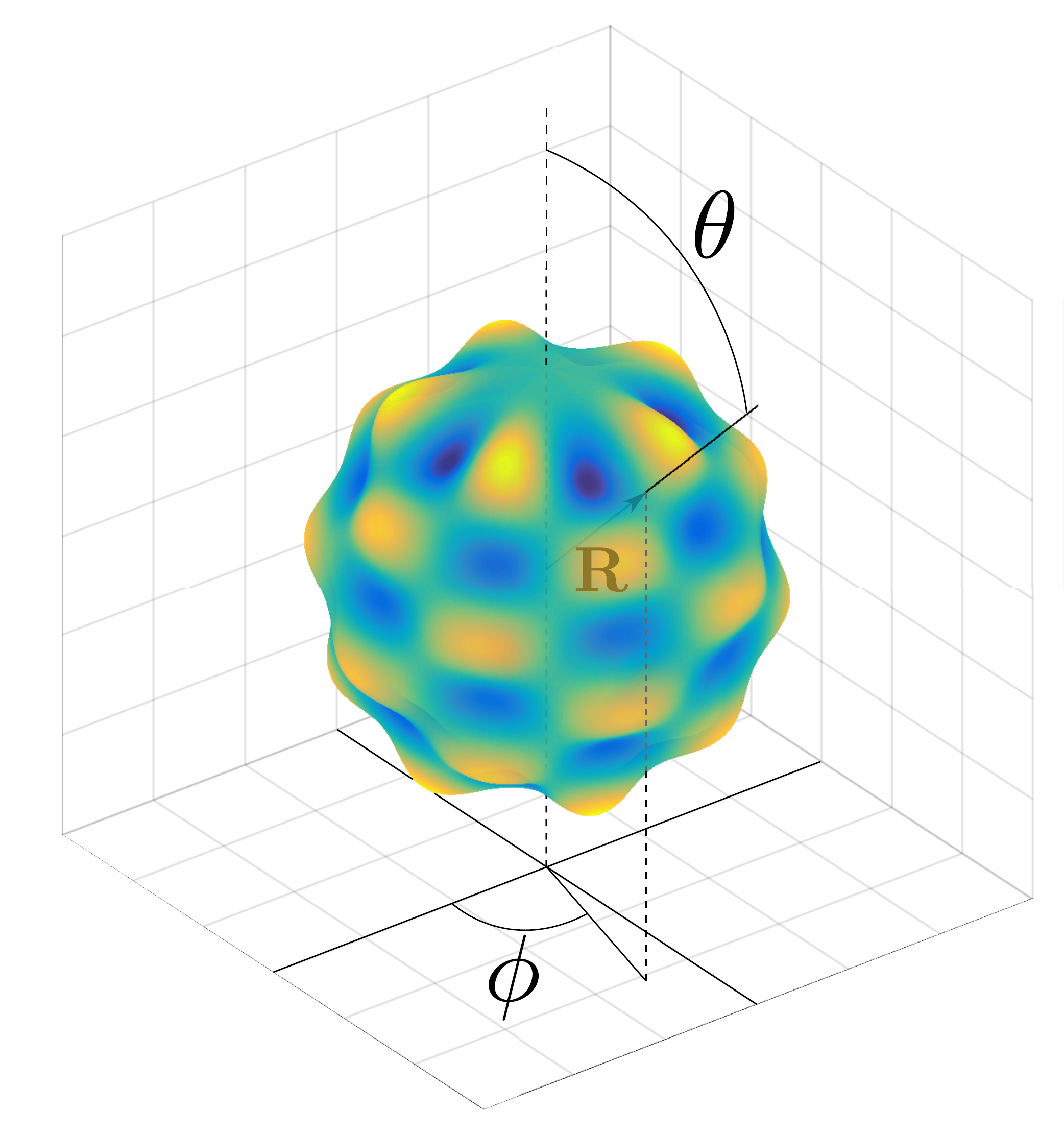}
\caption{The shape of the almost spherical membrane is described by a vector function $\mathbf{R}(\theta,\phi)$, whereas the protein distribution is described by a scalar function $\rho(\theta,\phi)$ represented by the colour-coding, e.g.~yellow and blue could correspond to high and low protein density, respectively.}
   \label{fig:geometry}
\end{figure}

We will adopt a continuum elastic model of a closed membrane, which might represent a model vesicle or a biological cell, and study the stability of spherical shapes to perturbations in the presence of curvature-inducing proteins that decorate the membrane. The shape of a quasi-spherical membrane can be written in spherical coordinates as $\mathbf{R}(\theta,\phi) = R[1+u(\theta,\phi)]\mathbf{\hat{r}}$, where $R$ is the radius of the unperturbed sphere, $u(\theta,\phi)$ is a scalar function that describes the deviations from the sphere, and $\mathbf{\hat{r}}$ is the radial unit vector, see figure~\ref{fig:geometry}. The distribution of proteins on the membrane can be described in a similar way, with the surface number density $\rho(\theta,\phi) = \rho_0 [ 1+\psi(\theta,\phi) ]$. Here, $\rho_0$ is the average protein number density, i.e. $\rho_0 = N / 4 \pi R^2$ if $N$ is the total number of proteins on the membrane, and the function $\psi(\theta,\phi)$ represents the deviations from a homogeneous distribution of proteins.

We will assume that each protein covers a patch of membrane of area $a_0$, and imposes a spontaneous curvature $C_\mathrm{p}$ on the membrane, see figure~\ref{fig:scheme}. The bending free energy of the membrane can then be written within the spontaneous curvature model \cite{helf73,leib86,seif91,rama00} as
\begin{equation}
F_\mathrm{b} = \frac{\kappa}{2} \int \mathrm{d}A \; [ C^2 - 2 C_\mathrm{p} \rho a_0 C ]
\label{eq:Fb}
\end{equation}
where $\kappa$ is the bending rigidity of the membrane, and $C$ is the local membrane curvature, with $C=C_1+C_2$, where $C_1$ and $C_2$ are the two principal curvatures. The second term inside the integral represents the simplest possible coupling between protein density and local curvature. It can also be interpreted as a position-dependent spontaneous curvature $C_0(\theta,\phi) \equiv C_\mathrm{p} \rho (\theta,\phi) a_0$, which varies from $C_0 = 0$ in the absence of proteins, with $\rho=0$, to $C_0 = C_\mathrm{p}$ for full coverage of proteins, with $\rho = 1 / a_0$.  The local membrane curvature $C(\theta, \phi)$ can be written explicitly as a function of $u(\theta, \phi)$, as described in ref~\citenum{helf86}.

Besides the bending contributions to the free energy, we need to take into account the entropic contributions due to the mixing and density fluctuations of the proteins. To lowest order, this contribution to the free energy can be incorporated as
\begin{equation}
F_\mathrm{d} = \frac{1}{2\chi} \int \mathrm{d}A \; [ \xi^2 (\nabla \psi)^2 + \psi^2]
\label{eq:Fd}
\end{equation}
Here, $\chi$ and $\xi$ are the compressibility and the correlation length of the protein density fluctuations, respectively. The first term in the integral penalises the creation of interfaces between high protein density and low protein density regions, whereas the second term penalises deviations from a homogeneous protein distribution.

We will also consider the effect of the tethering of the membrane to a cell wall or actomyosin cortex, by including a harmonic confinement potential of the form
\begin{equation}
F_\mathrm{h} = \frac{k_\mathrm{te}R^2}{2} \int \mathrm{d}A \; u^2
\label{eq:Fh}
\end{equation}
where $k_\mathrm{te}$ is an effective spring constant per unit area,  which in general may include contributions from specific interactions (i.e. proteins that directly link the membrane to the wall/cortex) as well as non-specific interactions such as steric repulsion, van der Waals attraction or electrostatic attraction/repulsion. Within this effective description, the cell wall/cortex is taken to be spherical and rigid (i.e. much more rigid than the membrane), and $k_\mathrm{te}$ penalises deviations of the membrane position from the (optimal) equilibrium membrane-wall distance.

Lastly, we consider the possibility that the membrane is connected to a membrane area reservoir at constant membrane tension. A constant membrane tension is typical of biological cells, \cite{morr01,sens15} and can be mimicked in model vesicle systems by the use of micropipette aspiration. The contribution of a membrane tension $\sigma$ to the free energy is
\begin{equation}
F_\mathrm{t} = \sigma \int \mathrm{d}A
\label{eq:Ft}
\end{equation}

The total free energy can finally be written as the sum of these four contributions, with
\begin{equation}
F = F_\mathrm{b} + F_\mathrm{d} + F_\mathrm{h} + F_\mathrm{t} = \int \mathrm{d}A \; \mathcal{F}
\label{eq:F}
\end{equation}
with the free energy density
\begin{equation}
\mathcal{F} \equiv \frac{\kappa}{2} [ C^2 - 2 C_\mathrm{p} \rho a_0 C ] + \frac{1}{2\chi} [ \xi^2 (\nabla \psi)^2 + \psi^2] + \frac{kR^2}{2} u^2 + \sigma
\label{eq:Fdens}
\end{equation}
 In addition, we will explicitly impose constraints on the volume enclosed by the membrane (representing osmotic balance), so that 
\begin{equation}
\frac{4 \pi R^3}{3} = \int \mathrm{d}V
\label{eq:Vconstr}
\end{equation}
as well as on the total number of proteins $N$ on the membrane, so that 
\begin{equation}
N = \rho_0 4 \pi R^2 = \int \mathrm{d}A \; \rho
\label{eq:Nconstr}
\end{equation}
at all times.

\subsection{Dynamics}

The effective force exerted on the membrane in the radial direction will be balanced by a frictional force, leading to a dynamical equation for the shape of the membrane as a function of time $t$
\begin{equation}
\partial_t u(\theta,\phi,t) = - L_u \frac{\delta F}{\delta u(\theta,\phi)}
\label{eq:dyna1}
\end{equation}
where $L_u$ is a transport coefficient corresponding to the membrane mobility.

On the other hand, the dynamical equation describing the diffusion of the proteins on the membrane can be written in the form of a continuity equation
\begin{equation}
\partial_t \psi(\theta,\phi,t) + \nabla \cdot \mathbf{J} = 0
\label{eq:dyna2a}
\end{equation}
with a current density $\mathbf{J} = - L_\psi \nabla \mu$, where $L_\psi$ is another transport coefficient and $\mu = \delta \mathcal{F} / \delta \psi(\theta,\phi)$ is the chemical potential. Putting all together, the dynamical equation for the protein density becomes
\begin{equation}
\partial_t \psi(\theta,\phi,t) =  L_\psi \nabla^2 \left( \frac{\delta F}{\delta \psi(\theta,\phi)} \right)
\label{eq:dyna2}
\end{equation}

The Laplacian operator on a sphere can be written as $\nabla^2 \equiv -\frac{1}{R^2} \hat{L}^2$, with the operator
\begin{equation}
-\hat{L}^2 \equiv \frac{1}{\sin \theta} \partial_\theta (\sin \theta \partial_\theta) + \frac{1}{\sin^2 \theta} \partial^2_\phi
\label{eq:L}
\end{equation}
This operator is diagonal in the basis of spherical harmonics $Y_{\ell m}(\theta,\phi)$. In particular, it satisfies
\begin{equation}
\hat{L}^2 Y_{\ell m}(\theta,\phi) = \ell (\ell+1) Y_{\ell m}(\theta,\phi)
\label{eq:Laction}
\end{equation}

\subsection{Linear stability analysis}

To leading order in $u$ and $\psi$, and taking into account the constraints (\ref{eq:Vconstr}--\ref{eq:Nconstr}) on the enclosed volume and total number of proteins on the membrane, we can write equations~(\ref{eq:dyna1}) and (\ref{eq:dyna2}) as
\begin{equation}
\partial_t u(\theta,\phi,t) = - L_u \left[ \left( \frac{\kappa}{R^2} \hat{L}^2 + \sigma \right) \left( \hat{L}^2 - 2 \right) u +  k_\mathrm{te} R^2 u + \frac{\kappa C_\mathrm{p}a_0 \rho_0}{R} \left( \hat{L}^2 - 2 \right) \psi  \right]
\label{eq:dyna1b}
\end{equation}
and
\begin{equation}
\partial_t \psi(\theta,\phi,t) = - \frac{L_\psi}{R^2} \left[ \frac{1}{\chi} \hat{L}^2 \psi + \frac{1}{\chi} \left( \frac{\xi^2}{R^2} \right) \hat{L}^4 \psi + \frac{\kappa C_\mathrm{p} a_0 \rho_0}{R} \hat{L}^2 \left( \hat{L}^2 - 2 \right) u \right]
\label{eq:dyna2b}
\end{equation}

We can write the solutions $u(\theta, \phi, t)$ and $\psi(\theta, \phi, t)$ as a sum of spherical harmonics, which provide a complete set of orthogonal functions on the sphere, so that
\begin{equation}
u(\theta,\phi,t) = \sum_{\ell,m} u_{\ell m}(t) Y_{\ell m}(\theta, \phi)~~\text{and}~~\psi(\theta,\phi,t) = \sum_{\ell,m} \psi_{\ell m}(t) Y_{\ell m}(\theta, \phi)
\label{eq:Y}
\end{equation}
where $u_{\ell m}$ and $\psi_{\ell m}$ are the amplitudes of the corresponding modes, and we have $\ell = 0, 1, 2...$ and $|m| \leq \ell$. However, the constraints (\ref{eq:Vconstr}--\ref{eq:Nconstr}) on the enclosed volume and total number of proteins on the membrane imply that the zero-amplitudes $u_{00}$ and $\psi_{00}$ cannot be varied independently. Explicitly imposing these constraints results in expressions for $u_{00}$ and $\psi_{00}$ as a function of the squared amplitudes of all modes
\begin{equation}
\sqrt{4\pi} u_{00} = - \sum_{\ell,m} u_{\ell m}^2
\label{eq:u00}
\end{equation}
\begin{equation}
\sqrt{4\pi} \psi_{00} = \frac{1}{2} \sum_{\ell,m} [2 - \ell(\ell+1)]u_{\ell m}^2 - 2 \sum_{\ell,m} u_{\ell m} \psi_{\ell m}
\label{eq:psi00}
\end{equation}
Equations (\ref{eq:u00}) and (\ref{eq:psi00}) imply that $u_{00}$ and $\psi_{00}$ are a function of the higher-order amplitudes, and furthermore, that they are of quadratic order (they are equal to a sum of $u_{\ell m}^2$ and $u_{\ell m} \psi_{\ell m}$ terms). For this reason, the $u_{00}$ and $\psi_{00}$ terms are negligible to linear order, and we can rewrite (\ref{eq:Y}) as
\begin{equation}
u(\theta,\phi,t) \simeq \sum_{\ell \geq 1,m} u_{\ell m}(t) Y_{\ell m}(\theta, \phi)~~\text{and}~~\psi(\theta,\phi,t) \simeq \sum_{\ell \geq 1,m} \psi_{\ell m}(t) Y_{\ell m}(\theta, \phi)
\label{eq:Y2}
\end{equation}

Inserting (\ref{eq:Y2}) into (\ref{eq:dyna1b}) and (\ref{eq:dyna2b}), we can rewrite the dynamical equations as separate equations for each of the $\ell \geq 1$ modes. Introducing a rescaled time variable 
\begin{equation}
\tau \equiv \left( \frac{\kappa L_u}{R^2} \right) t
\label{eq:tau}
\end{equation}
as well as dimensionless parameters
\begin{equation}
K \equiv \frac{k_\mathrm{te} R^4}{\kappa},~T \equiv \frac{\sigma R^2}{\kappa},~P \equiv \frac{\xi^2}{R^2},~M \equiv \frac{L_\psi}{\kappa L_u \chi},~S \equiv \rho_0 a_0 C_\mathrm{p} R,~\text{and}~B \equiv \frac{\kappa \chi}{R^2}
\label{eq:dimless}
\end{equation}
the equations become
\begin{equation}
- \partial_\tau u_{\ell m} = \left\{ \left[ \ell (\ell+1) + T \right] (\ell+2) (\ell-1) + K \right\} u_{\ell m} + S  (\ell+2) (\ell-1) \psi_{\ell m}
\label{eq:dyna1c}
\end{equation}
and
\begin{equation}
- (1/M) \partial_\tau \psi_{\ell m} = B S \ell (\ell+1)  (\ell+2) (\ell-1) u_{\ell m} +  \ell (\ell+1) \left[ 1 + P \ell (\ell+1) \right] \psi_{\ell m}
\label{eq:dyna2c}
\end{equation}

The solutions to (\ref{eq:dyna1c}) and (\ref{eq:dyna2c}) will have the form
\begin{equation}
u_{\ell m}(\tau) = u_{\ell m}(0) \mathrm{e}^{\lambda \tau},~~\psi_{\ell m}(\tau) = \psi_{\ell m}(0) \mathrm{e}^{\lambda \tau}
\label{eq:u_psi}
\end{equation}
Inserting these solutions back into (\ref{eq:dyna1c}) and (\ref{eq:dyna2c}), and setting the determinant of the coefficients to zero, we can obtain an equation for the growth rates $\lambda$ of the characteristic modes of the system, which reads
\begin{equation}
\lambda^2 + b \lambda + c = 0
\label{eq:modes}
\end{equation}
with coefficients
\begin{equation}
b \equiv  K - 2T +  \ell (\ell+1) \left[ M + T - 2 + \ell (\ell+1) (MP+1) \right]
\label{eq:defb}
\end{equation}
and
\begin{equation}
c \equiv M \ell (\ell+1)  \left\{ \left[ K + \left( \ell (\ell+1) + T \right) (\ell+2) (\ell-1) \right] \left[1+P \ell (\ell+1) \right] - W (\ell+2)^2 (\ell-1)^2 \right\}
\label{eq:defc}
\end{equation}
where we have defined the parameter
\begin{equation}
W \equiv BS^2 = \kappa \chi \rho_0^2 a_0^2 C_\mathrm{p}^2
\label{eq:defW}
\end{equation}
The two characteristic modes of the system given by the solutions to (\ref{eq:modes}) can finally be written as
\begin{equation}
\begin{split}
\lambda_\pm &=  \frac{1}{2} \Big[ -K + 2T - \ell (\ell+1) \left[ M + T - 2 + \ell (\ell+1) (MP+1) \right] \Big]  \\ 
&\pm \frac{1}{2} \sqrt{\Big[ K - 2T - \ell (\ell+1) \left[ M - T + 2 + \ell (\ell+1) (MP-1) \right] \Big]^2 + 4MW \ell(\ell+1) (\ell+2)^2 (\ell-1)^2 } 
\end{split}
\label{eq:lambda}
\end{equation}

Because $b$ in (\ref{eq:modes}) always satisfies $b>0$ for all modes with $\ell \geq 1$, we know that the amplitude with the smaller value, $\lambda_-$, is always negative for all $\ell$-modes. On the other hand, the larger one, $\lambda_+$, might be positive or negative depending on the $\ell$-mode and on the values of the parameters $W$, $K$, $T$, $P$, and $M$.  It is also worth noting that the stability analysis is independent of the value of $m$ of the spherical harmonics. This ultimately arises from the fact that the eigenvalues of the Laplacian of a spherical harmonic are independent of its $m$-value.

The physical significance of the five dimensionless parameters is the following. The parameter $W$ represents the protein-induced spontaneous curvature, and increases both with the average density $\rho_0$ of proteins on the membrane and with the characteristic spontaneous curvature $C_\mathrm{p}$ of these proteins. The parameter $K$ represents the strength of the confinement of the membrane by its interaction with the rigid cell wall/cortex. The parameter $T$ represents the magnitude of the membrane tension. The parameter $P$ compares the correlation length of the protein density fluctuations to the size of the cell or vesicle. Given that correlation lengths are typically of the order of nanometers whereas cell or vesicle sizes are of the order of micrometers, $P$ will generally be small, and will decrease or increase with increasing or decreasing cell/vesicle size, respectively. Finally, the parameter $M$ compares the typical timescale of the changes in membrane shape ($R^2 / L_u \kappa$) to that of changes in protein distribution ($\chi R^2 / L_\psi$). Importantly, we note that all five dimensionless parameters are always positive.

\section{Results \label{sec:results}}

A positive value of the mode amplitude $\lambda_+$ implies that fluctuations of this mode will grow instead of decaying, and therefore modes with $\lambda_+>0$ are unstable. If, by small changes in one of the system parameters $W$, $K$, $T$, $P$, or $M$, one of the modes $\lambda_+$ switches from having a negative value to having a positive value, the system will exhibit spontaneous pattern formation. In the following, we will explore the conditions under which spontaneous pattern formation occurs.

First of all, we note that, as described above, the $\ell=0$ mode cannot vary independently as it is fixed by the constraints on the enclosed volume and total number of proteins, see (\ref{eq:u00}--\ref{eq:psi00}). Furthermore, by substituting $\ell=1$ in (\ref{eq:lambda}), we find the mode amplitudes $-K$ and $-2M(1+2P)$, which can never be positive, implying that the $\ell=1$ mode can never become unstable. It can, however, become marginally stable in the particular case of $K=0$, i.e.~in the absence of tethering to the cell wall. This reflects the fact that $\ell = 1$ deformations of the membrane shape are equivalent to spatial translations, and that the curvature energy of the membrane is invariant to such translations. The presence of the cell wall, however, breaks translational invariance. All things considered, instabilities and therefore spontaneous pattern formation can occur only for higher modes $\ell \geq 2$, which we will discuss below. 

The larger solution $\lambda_+$ of (\ref{eq:modes}) will be positive, with $\lambda_+>0$, if and only if $c<0$. Using the definition of $c$ in (\ref{eq:defc}), this condition can be rewritten as
\begin{equation}
W > \frac{\{ K + [\ell (\ell+1)+T] (\ell+2) (\ell-1) \} [1 + P \ell (\ell+1)]}{  (\ell+2)^2 (\ell-1)^2 } \equiv W_\ell
\label{eq:Wcrit}
\end{equation}
which serves as a definition of $W_\ell$, the critical value of the parameter $W$ above which mode $\ell$ becomes unstable. Going back to the definition of $W$ in (\ref{eq:defW}), the inequality (\ref{eq:Wcrit}) implies that an increase in the average density $\rho_0$ of curvature-inducing proteins beyond a critical density will trigger an instability with spherical harmonic mode $\ell$ in both the shape and protein distribution of the membrane. Furthermore, the critical protein density that is needed to trigger an instability decreases with increasing protein spontaneous curvature $C_\mathrm{p}$. Importantly, we note that the critical value $W_\ell$ is independent of the parameter $M$, and therefore depends only on three parameters, $P$, $K$, and $T$. In fact, the parameter $M$ drops out of all relevant equations in the following, so that pattern formation in the system turns out to be governed by only four dimensionless parameters: $W$, $K$, $T$, and $P$. This is a consequence of the fact that $M$ is a mobility parameter that related the timescale of changes in membrane shape to that of changes in protein distribution, and as such it only affects  the dynamics of the system.

Alternatively, the instability condition $c<0$ can be written as
\begin{equation}
K < \frac{W  (\ell+2)^2 (\ell-1)^2  }{1+P\ell(\ell+1)} - [\ell (\ell+1) + T] (\ell+2) (\ell-1) \equiv K_\ell
\label{eq:Kcrit}
\end{equation}
or
\begin{equation}
T < \frac{W (\ell+2) (\ell-1)}{1+P\ell(\ell+1)} - \frac{K}{(\ell+2)(\ell-1)} - \ell (\ell+1) \equiv T_\ell
\label{eq:Tcrit}
\end{equation}
which define $K_\ell$ and $T_\ell$, the critical values of $K$ and $T$, respectively, below which mode $\ell$ becomes unstable. Going back to the definitions of $K$ and $T$ in (\ref{eq:dimless}), the inequalities (\ref{eq:Kcrit}) and (\ref{eq:Tcrit}) respectively imply that the shape and protein distribution instability can also be triggered by a decrease in the tethering strength of the membrane to the cell wall/cortex, or by a decrease in the membrane tension. Once again, we note that the critical values $K_\ell$ and $T_\ell$ are independent of the parameter $M$.

As outlined in the previous two paragraphs, the parameters that could presumably be actively controlled by a biological cell or tuned in experiments with model vesicles are $W$, i.e. the density of proteins on the cell surface, $K$, i.e. the tethering strength of the membrane to the cell wall/cortex, and $T$, the membrane tension. The parameter $P$, on the other hand, represents the correlation length of the protein density fluctuations, i.e. the typical distance at which proteins can sense each other, and will in general be fixed for a given system. It therefore makes sense to explore the behaviour of the system when $W$, $K$, and $T$ are varied for a fixed value of $P$.

\begin{figure}[h]
\centering
\includegraphics[width=1\linewidth]{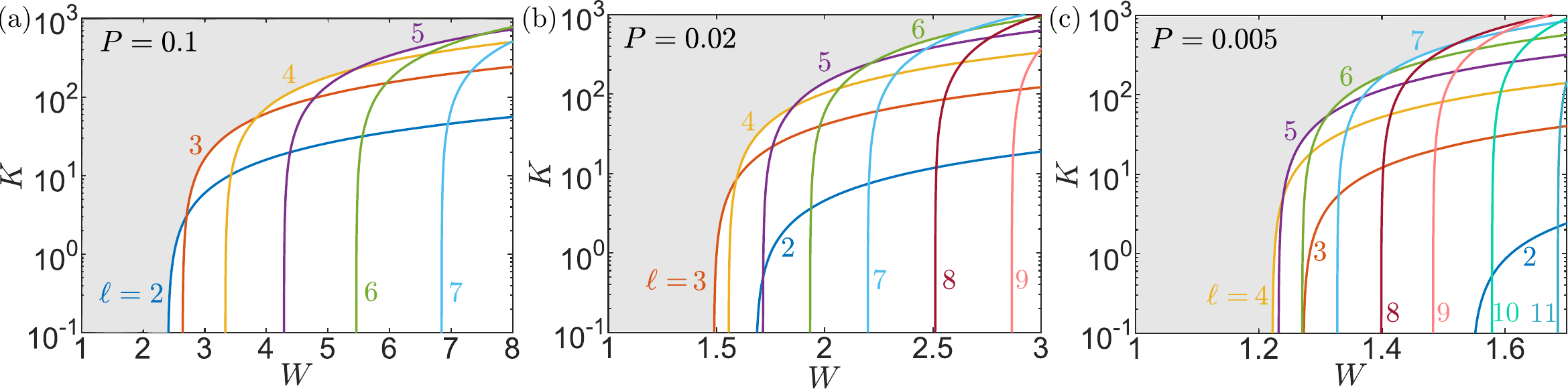}
\caption{  Instability lines $W=W_\ell(K)$ for modes $\ell \geq 2 $, vanishing tension $T=0$, and three values of $P$: (a) $P=0.1$, (b) $P=0.02$, and (c) $P=0.005$. For low $W$ (i.e. low number of curvature-inducing proteins) and high $K$ (i.e.~strong confinement of the membrane due to tethering to the cell wall/cortex), the spherical homogeneous state is stable (grey region). As $W$ is increased or $K$ is decreased, the system will hit an instability with a given value of $\ell$. If $W$ or $K$ keep increasing or decreasing, respectively, they will hit the instabilities of further modes. The higher the value of $K$ or $W$, the higher the value of $\ell$ of the first unstable mode as $W$ is increased or $K$ is decreased. The parameter $P$ represents the correlation length of the protein density fluctuations.}
   \label{fig:WvsK}
\end{figure}

Using (\ref{eq:Wcrit}), in figure~\ref{fig:WvsK} we have plotted the lines $W=W_\ell(K)$ for $\ell \geq 2$, using $T=0$ (i.e.~negligible membrane tension) and three different values of $P$, namely $P=0.1$, $0.02$, and $0.005$. For a vesicle/cell of radius $1~\mu$m, these values of $P$ would correspond to correlation lengths of $\xi = 320$~nm, 140~nm, and 70~nm, respectively. In the region of low $W$ and high $K$, depicted in grey, the spherical state with a homogeneous protein distribution is stable. As $W$ is increased from low values, the system will hit the instability of the first unstable mode, with a given value of $\ell$ which will depend on the value of $K$. Alternatively, if $K$ is decreased from high values, the system will also hit the instability of the first unstable mode with a given $\ell$ which will depend on the value of $W$. The higher the value of $K$, the higher the value of $\ell$ of the first unstable mode as $W$ is increased. Similarly, the higher the value of $W$, the higher the value of $\ell$ of the first unstable mode as $K$ is decreased.

There are important differences in the way in which $W$ and $K$ act to trigger pattern formation. Independently of the value of $K$, and even for $K=0$, a sufficiently high $W$ will always lead to pattern formation. On the other hand, a decrease in $K$ can only lead to pattern formation if $W$ is above the critical value $W_\ell (K=0)$. Furthermore, we note that figure~\ref{fig:WvsK} has a semilogarithmic axis: whereas the critical value of $W$ above which pattern formation occurs is always in the vicinity of 1, with $W \gtrsim 1$, the critical value of $K$ below which pattern formation occurs can vary over many orders of magnitude. Pattern formation is therefore particularly sensitive to $W$,~i.e. to the density of curvature-inducing proteins on the membrane.

And what is the effect of $P$, that is, of the correlation length of the protein density fluctuations? Let us now compare figures~\ref{fig:WvsK}(a), (b), and (c). For the highest value of $P$, in (a), the first unstable mode for increasing $W$ at vanishing $K$ is $\ell=2$, whereas larger values of $K$ lead to the instabilities of higher-order modes with $\ell > 2$. As $P$ is decreased, as in (b), the first unstable mode at vanishing $K$ is now $\ell=3$: the mode $\ell = 2$ is not the first unstable mode for any value of $K$. When $P$ is decreased even further, as in (c), $\ell=4$ becomes the first unstable mode at vanishing $K$, and neither $\ell=2$ nor $\ell=3$ are the first unstable modes for any value of $K$. This trend continues as $P$ is decreased further, with progressively higher order modes becoming the first unstable mode at vanishing $K$. Moreover, we note that, as $P$ is decreased, the critical value of $W$ above which pattern formation occurs moves closer and closer to $W=1$.

In figure~\ref{fig:WvsK} we have explored the stability behaviour of the system as a function of $W$ and $K$, for fixed vanishing tension $T=0$. Considering a fixed non-zero tension $T>0$ leads to the same qualitative behaviour of the system as a function of $W$ and $K$. Furthermore, the behaviour of the system as a function of $W$ and $T$ for fixed $K$ is qualitatively identical to that as a function of $W$ and $K$ for fixed $T$, leading to instability lines analogous to those in figure~\ref{fig:WvsK}. We thus omit these results for the sake of brevity.

\begin{figure}[h]
\centering
\includegraphics[width=1\linewidth]{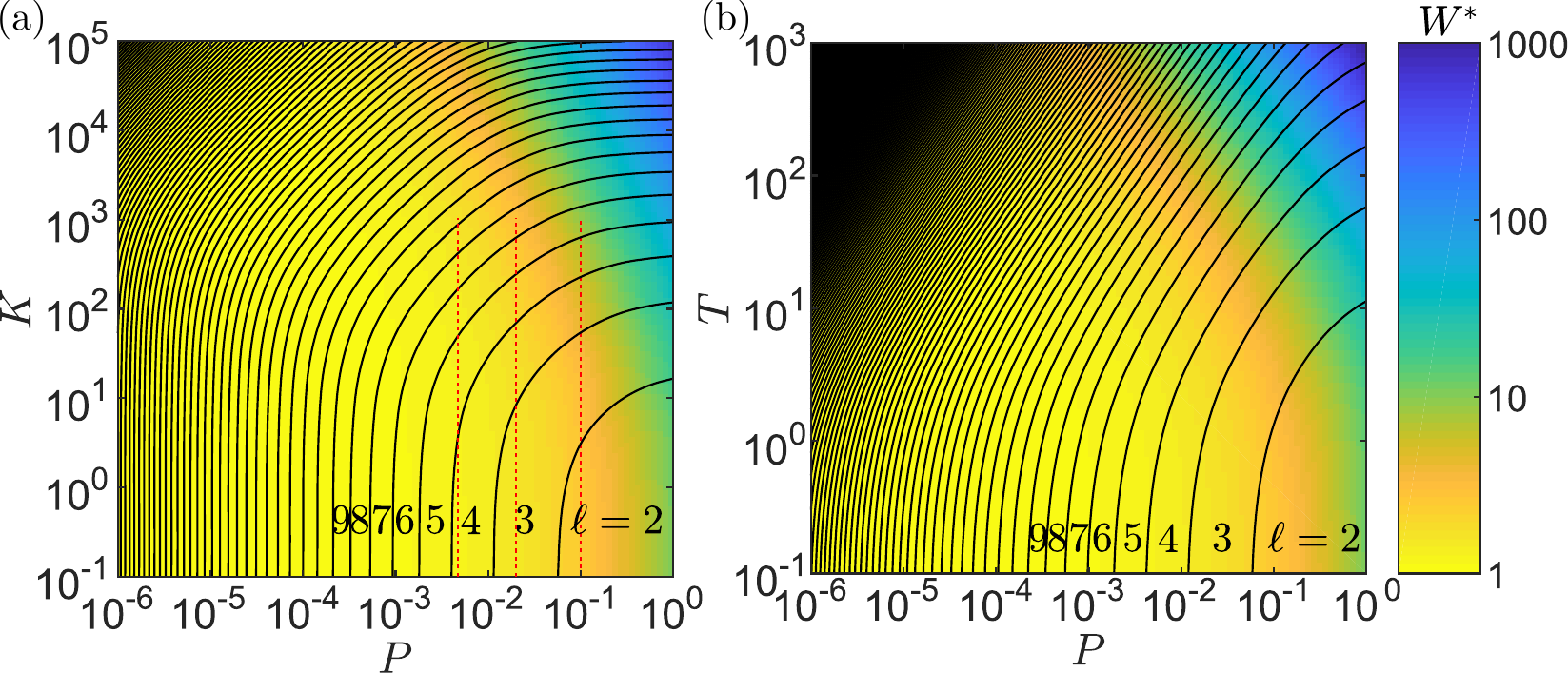}
\caption{  Pattern formation triggered by an increase in the number of curvature-inducing proteins: First unstable modes $\ell^*_W$ when $W$ is increased, (a) as a function of the parameters $P$ and $K$ for $T=0$; and (b) as a function of the parameters $P$ and $T$ for $K=0$. The critical value $W^*$ above which pattern formation occurs can be calculated from (\ref{eq:Wcrit}), and is indicated by the colour-coding, which is the same for (a) and (b). The vertical dashed lines in (a) correspond to the three particular cases $P=0.1$, $P=0.02$ and $P=0.005$ displayed in figure~\ref{fig:WvsK}.  }
   \label{fig:PvsK}
\end{figure}

As just described, in order to characterise the system, it is particularly important to identify the first unstable mode when $W$ is increased, that is, the mode with smallest $W_\ell$ for given values of $P$, $K$, and $T$, which we will denote as $\ell^*_W$. The critical value of $W$ above which the first unstable mode becomes unstable is then $W^* \equiv W_{\ell^*_W} \equiv \min_\ell (W_\ell)$.  The boundaries between the regions in the three-dimensional $(P,K,T)$ parameter space in which modes $\ell$ and $\ell+1$ are the first unstable mode for increasing $W$ can be obtained from the condition $W_\ell = W_{\ell+1}$, which can be written explicitly using (\ref{eq:Wcrit}) as
\begin{equation}
P = \frac{(\ell-1)\ell(\ell+2)(\ell+3)(2+T)-2K[1-\ell(\ell+2)]}{(\ell-1)\ell(\ell+2)(\ell+3)\{[\ell(\ell+2)-4](\ell+1)^2-2T\}-K[\ell(\ell+1)^2 (\ell+2)-4]}
\label{eq:Klines}
\end{equation}
In figure~\ref{fig:PvsK}, we have used equation~(\ref{eq:Klines}) to explore pattern formation in (a) the $(P,K,T=0)$ plane and (b) the $(P,K=0,T)$ plane. For any point in $(P,K,T)$ space, we can obtain the critical value $W^*$ above which pattern formation occurs, using (\ref{eq:Wcrit}). This information is also colour-coded in figure~\ref{fig:PvsK}. Several important observations can be made: (i) Once again, we see that $K$ and $T$ have qualitatively similar effects in pattern formation. (ii) Both an increase in $K$ or $T$, as well as a decrease in $P$ lead to increasingly higher-order modes being the first unstable mode. (iii) In most regions of the parameter space, the critical value $W^*$ above which pattern formation occurs is very close to 1. The only exception is the region of $P \approx 1$ and large $K$ or $T$, in which $W^*$ can be much larger than one.

A particularly important case, with regards to its experimental relevance, is that of a model lipid vesicle, for which we have both $K=0$ (there is no wall or cortex attached to the membrane) and $T=0$ (if we are considering a flaccid, unstretched vesicle). This corresponds to the bottom part of of both figure~\ref{fig:PvsK}(a) and (b). In this limit case, which mode first becomes unstable when $W$ (i.e.~the number of curvature-inducing proteins on the membrane) is increased depends only on the parameter $P$ (i.e.~the correlation length of the protein density fluctuations), with the boundaries between $\ell$ and $\ell+1$ being the first unstable modes given by the simple expression
\begin{equation}
P = \frac{2}{[\ell(\ell+2)-4](\ell+1)^2}
\label{eq:Klines0}
\end{equation}
as obtained from equation~(\ref{eq:Klines}) with $K=T=0$. Using equation~(\ref{eq:Klines0}), we predict that for a tensionless spherical vesicle, the $\ell=2$ mode will be the first unstable mode if $P>1/18$, the $\ell=3$ mode will be the first unstable mode if $1/18>P>1/88$, the $\ell=4$ mode if $1/88>P>1/250$, and so on. For a typical vesicle of radius $1~\mu$m and a typical correlation length of $\xi = 20$~nm, we have $P=4 \cdot 10^{-4}$, and we find that the $\ell=8$ mode will be the first unstable mode.

\begin{figure}[h]
\centering
\includegraphics[width=1\linewidth]{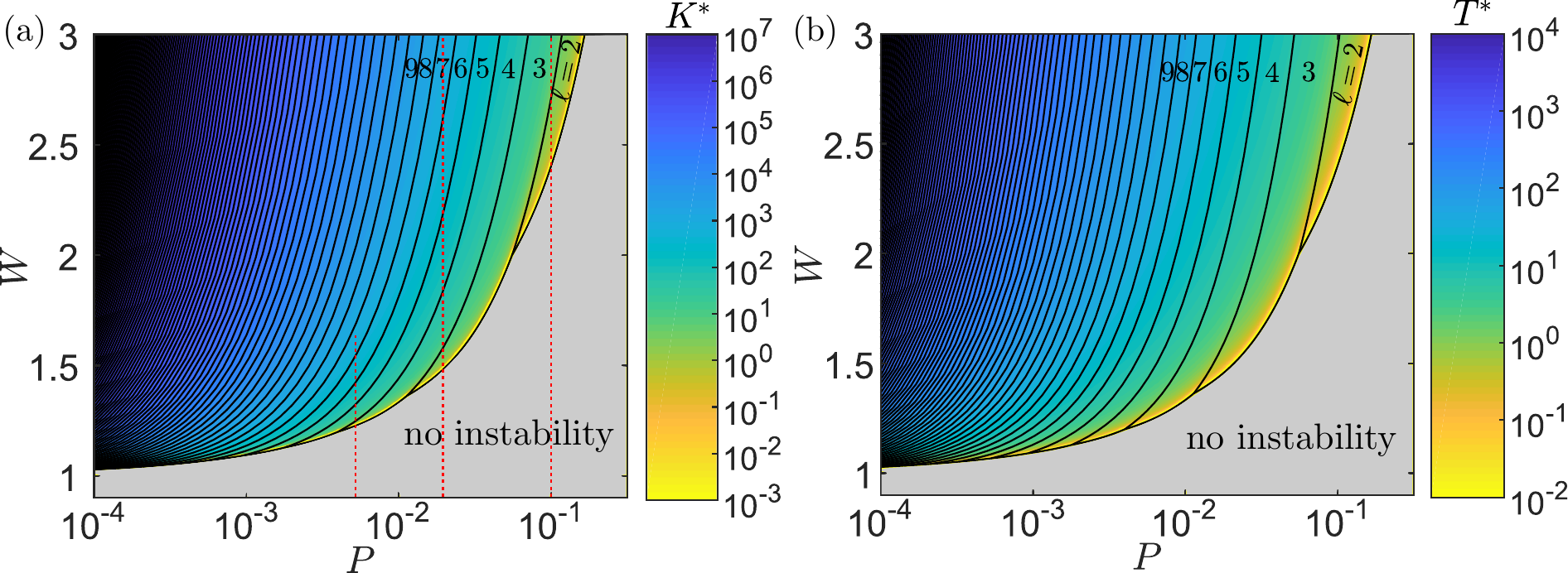}
\caption{  (a) Pattern formation triggered by a decrease in the tethering strength of the membrane to the cell wall/cortex: First unstable modes $\ell^*_K$ when $K$ is decreased, as a function of the parameters $P$ and $W$ for $T=0$. The critical value $K^*$ below which pattern formation occurs can be calculated from (\ref{eq:Kcrit}), and is indicated by the colour-code. The vertical dashed lines correspond to the three particular cases $P=0.1$, $P=0.02$ and $P=0.005$ displayed in figure~\ref{fig:WvsK}. (b) Pattern formation triggered by a decrease in membrane tension: First unstable modes $\ell^*_T$ when $T$ is decreased, as a function of the parameters $P$ and $W$ for $K=0$. The critical value $T^*$ below which pattern formation occurs can be calculated from (\ref{eq:Tcrit}), and is indicated by the colour-code. }
   \label{fig:PvsW}
\end{figure}

Alternatively, we could ask ourselves what is the first unstable mode $\ell^*_K$ when $K$ is decreased for given values of $P$, $W$, and $T$, or equivalently, the mode with largest $K_\ell$ for given $P$, $W$, and $T$. The critical value of $K$ below which the first unstable mode becomes unstable is then $K^* \equiv K_{\ell^*_K} \equiv \max_\ell (K_\ell)$.  The boundaries between the regions in the three-dimensional $(P,W,T)$ parameter space in which modes $\ell$ and $\ell+1$ are the first unstable mode for decreasing $K$ can be obtained from the condition $K_\ell = K_{\ell+1}$, which can be written explicitly using (\ref{eq:Kcrit}). The resulting $(P,W)$ stability diagram for the particular case of $T=0$ is shown in figure~\ref{fig:PvsW}(a). In the same way, we can find the first unstable mode $\ell^*_T$ when $T$ is decreased for given values of $P$, $W$, and $K$, with a critical value given by $T^* \equiv T_{\ell^*_T} \equiv \max_\ell (T_\ell)$, and boundaries in the $(P,W,K)$ parameter space given by $T_\ell = T_{\ell+1}$, which can be written explicitly using (\ref{eq:Tcrit}). The resulting $(P,W)$ stability diagram for the particular case of $K=0$ is shown in figure~\ref{fig:PvsW}(b).

Once again, we find that $K$ (the strength of the tethering of the membrane to the cell wall/cortex) and $T$ (the membrane tension) behave in a qualitatively similar way. As expected from figure~\ref{fig:WvsK}, an instability can only occur for decreasing $K$ (or $T$) if $W$ is sufficiently high. This minimum value of $W$ required for pattern formation approaches $W=1$ for small $P$. Indeed, figure~\ref{fig:PvsW} illustrates very clearly a striking feature of the system: for low values of $P$ (which are the most typical given that the correlation length of protein density fluctuations $\xi$ is normally much smaller than the membrane radius $R$), values of $W$ only slightly above 1 can lead to the instability of modes with very high $\ell$ when $K$ or $T$ are decreased. This is evidenced by the high density of boundary lines in the region of low $P \ll 1$, $W \gtrsim 1$.

\section{Discussion \label{sec:discussion}}

\subsection{  Estimation and control of model parameters in real systems  }

We have shown above that pattern formation in a spherical membrane containing curvature-inducing proteins is controlled by the four dimensionless parameters $W$, $K$, $T$ and $P$, which represent the number of curvature-inducing proteins on the membrane, the strength of the membrane tethering to the cell wall/cortex, the membrane tension, and the correlation length of protein fluctuations, respectively. An important question is then: what are the typical values of these parameters in real systems, and to what extent can they be controlled by a biological cell, or tuned in experiments with model vesicles?

The parameter to which the system is most sensitive is $W$, see figures~\ref{fig:WvsK}, \ref{fig:PvsK} and \ref{fig:PvsW}. Even if $K$, $T$, and $P$ vary across many orders of magnitude, the critical value $W$ above which pattern formation occurs always stays in the proximity of $W^* \approx 1$, except in the extreme case of very high $K$ (or $T$) \emph{and} $P \approx 1$ simultaneously, see figure~\ref{fig:PvsK}. Going back to the definition of $W$ in terms of the dimensionful system parameters in (\ref{eq:defW}), we see that the requirement $W \geq W^* \approx 1$ implies that $\kappa \chi \rho_0^2 a_0^2 C_\mathrm{p}^2 \gtrsim 1$. Here, $\kappa$ is the bending rigidity of the membrane, $\chi$ is the compressibility of the protein density fluctuations, $\rho_0$ is the average density of curvature-inducing proteins on the membrane, $a_0$ is the lateral area of a single protein, and $C_\mathrm{p}$ is the protein spontaneous curvature. Furthermore, in the limit of low protein density $\rho_0 a_0 \ll 1$, the compressibility $\chi$ can be approximated by $\chi \approx 1/(k_\mathrm{B}T \rho_0)$, \cite{cai95} where $k_\mathrm{B}$ is Boltzmann's constant and $T$ is the temperature. In this limit, the requirement for pattern formation thus becomes $(\kappa/k_\mathrm{B}T)\rho_0 a_0^2 C_\mathrm{p}^2 \gtrsim 1$. In general, the protein spontaneous curvature $C_\mathrm{p}$ will be of the order of the (inverse) characteristic length of the protein $\sqrt{a_0}$, so that we can take $C_\mathrm{p}^2 a_0 \approx 1$. We finally conclude that pattern formation typically occurs for average protein densities satisfying
\begin{equation}
\rho_0 a_0 \gtrsim k_\mathrm{B}T / \kappa
\end{equation}
Here, $\rho_0 a_0$ is simply the dimensionless area fraction of membrane covered by the protein. Typical values of the bending rigidity of membranes range from $10$ to $100~k_\mathrm{B}T$, leading to a critical protein coverage of the order of $\rho_0 a_0 \sim 0.01$--$0.1$. Importantly, the range obtained self-consistently validates the low protein density assumption made above. Furthermore, such coverages are within the range achievable both in biological cells as well as in model vesicles. In this picture, a biological cell could up- or down-regulate the expression of the curvature-inducing protein in order to switch between patterned and non-patterned conformations, see figure~\ref{fig:PvsK}. Furthermore, the concentration of curvature-inducing proteins on the membrane could be directly controlled in experiments with model vesicles.

Let us now turn to the dimensionless parameters $K \equiv k_\mathrm{te}R^4 / \kappa$ and $T \equiv \sigma R^2 / \kappa$, which both act as geometric constraints on the membrane: $K$ represents the confinement of the membrane due to its interaction/tethering to the cell wall or cortex, whereas $T$ represents the membrane tension, which acts to minimise the cell membrane area. It is interesting to note that, while model membranes such as Giant Unilamellar Vesicles show clear shape fluctuations due to thermal excitation of bending modes, \cite{dimo06} eukaryotic cells or bacteria do not show such fluctuations. The latter is an indication that, in such systems, membrane confinement and tension must overpower bending, and consequently that in these systems $K \gg 1$ and/or $T \gg 1$, as can be confirmed by the quantitative estimates that follow.

Estimates of the confinement strength $k_\mathrm{te}$ of biological membranes due to the interaction with the corresponding cell wall or cortex do not abound in the literature. In Ref.~\citenum{aler15}, the density of membrane-cortex linkers in eukaryotic cells was estimated to be around $\rho_\mathrm{link} \approx 100~\mu$m$^{-2}$, whereas the spring constant of a typical linker was estimated to be $k_\mathrm{link} \approx 10^{-4}$~N/m. The effective tethering strength should go as $k_\mathrm{te} \approx \rho_\mathrm{link} k_\mathrm{link}$, leading to the estimate $k_\mathrm{te} \approx 10^{10}$~J/m$^4 \approx 2.5 \cdot 10^6~k_\mathrm{B}T/\mu$m$^4$. Considering a typical range of bending rigidities $\kappa = 10$--$100~k_\mathrm{B}T$, and a typical cell radius ranging from from $R=1~\mu$m to $10~\mu$m, we find values for $K$ ranging from $2.5 \cdot 10^4$ up to $2.5 \cdot 10^9$. Cells could then actively switch between patterned and non-patterned conformations by down- or up-regulating the concentration of linker proteins between the plasma membrane and the cell wall/cortex, see figure~\ref{fig:PvsW}(a).

The typical tension of cellular membranes, on the other hand, has been extensively measured for different cell types, and can range from $\sigma = 3$~pN/$\mu$m for epithelial cells up to about $\sigma = 300$~pN/$\mu$m for keratocytes. \cite{morr01,sens15} Using the range $\sigma = 3$--$300$~pN/$\mu$m for the membrane tension, together with the estimates $\kappa = 10$--$100~k_\mathrm{B}T$ for the bending rigidity of the membrane and $R=1$--$10~\mu$m for a typical cell radius, we obtain values of $T$ ranging from $7$ to $7 \cdot 10^5$. Cells can actively regulate their own tension in order to maintain homeostasis. \cite{morr01} In this way, cells could switch between patterned and non-patterned conformations by actively decreasing or increasing the tension of their plasma membrane, see figure~\ref{fig:PvsW}(b). Furthermore, triggering of pattern formation \emph{via} a decrease in membrane tension could be explored in experiments using model Giant Unilamellar Vesicles aspirated by micropipettes, which allows direct experimental control over the membrane tension.

Let us finally examine the dimensionless parameter $P$, defined as $P \equiv \xi^2/R^2$, where $\xi$ is the correlation length of the protein density fluctuations and $R$ is the radius of the cell or vesicle. The correlation length $\xi$ is a measure of the distance at which proteins or protein clusters can sense each other, typically \emph{via} membrane-mediated interactions in the absence of other long-ranged interactions. Previous work \cite{gole96a,gole96b,weik98,reyn11} has shown that the typical length scale of membrane-mediated interactions is the size of the curvature-inducing element itself, so that we can use an estimate of $\xi = 10$--$20$~nm. On the other hand, the radius of cells or cellular compartments, as well as of model vesicles, can range between $R = 100$~nm and $10~\mu$m. With this, we find a range of $P \sim 10^{-6}$--$10^{-2}$. In this range of values with $P \ll 1$, as described above, pattern formation is tighly controlled by the number of curvature-inducing proteins, with an instability occurring as soon as $W \gtrsim 1$. Moreover, the value of $P$ directly controls the $\ell$-order of the first unstable during pattern formation, and as a consequence controls the typical size of the protein-rich, highly-curved domains. The consequences of this fact are discussed in the following section.

\subsection{Biological relevance}

\subsubsection{Cell division}

Cell division requires polarisation of the cell, so that the spherical symmetry of the cell is broken, leading to two identifiable poles as well as an equatorial line. Spontaneous pattern formation \emph{via} an instability due to the presence of curvature-inducing proteins, as described here, provides a simple mechanism for such a symmetry breaking. This occurs for the mode $\ell=2$ of the instability, see figure~\ref{fig:applications}(a), which can be the first unstable mode as long as $P>1/18 \simeq 0.056$, as determined from (\ref{eq:Klines0}) and displayed on figures~\ref{fig:PvsK} and \ref{fig:PvsW}. For a typical cell size of $R=1~\mu$m, such values of $P$ would correspond to a correlation length $\xi$ for protein density fluctuations larger than 230~nm. This value appears too high for a typical protein, given that the correlation length is expected to be of the order of the protein size, i.e.~a couple of tens of nanometers. It could, however, be a plausible value for protein \emph{clusters}, composed of a few tens of proteins with lateral sizes of the order of 100~nm. If such clusters arose by a separate mechanism, a curvature-instability such as the one described here could lead to cell polarisation.

Several proteins, many of them related to cell division, are known to preferentially localise at the poles of bacterial membranes. \cite{shap02,lai04,thie07,bowm08,eber08} We note, however, that the generic mechanism proposed here is distinct and unrelated to the well-studied Min system, which serves to localise the FtsZ protein ring in rod-shaped bacteria. \cite{schw12,petr15} The Min system involves both membrane bound as well as cytosolic components, and locates the bacterial equator \emph{via} an oscillatory mechanism. The mechanism proposed here might however explain why FtsZ proteins can spontaneously self-assemble on vesicles, even in the absence of the Min system. \cite{osaw08,shlo09} In eukaryotic cells, cell division is mediated by the cytoskeleton, in particular by the mitotic spindle and the cleavage furrow. The mechanism underlying the initial positioning of this cell division apparatus is however not well understood at the molecular level. \cite{glot04,barr07}

Even if the mechanism described here did not play a direct role in cell division, it provides a generic pathway for symmetry breaking and the initiation of division of spherical membranes into two equally sized daughters, using only a minimal number of ingredients. As such, it could serve as a plausible mechanism for the division of protocells, as well as of synthetic cells in bottom-up synthetic biology. \cite{hanc03,zhu09,zwic16}

\subsubsection{Large-scale protein organisation}

Going beyond $\ell=2$, the mechanism described here also predicts pattern formation with modes of intermediate $\ell$, e.g.~in the range $\ell = 5$--$10$. Formation of such patterns would imply protein-rich, strongly-curved clusters with sizes on the order of $1/5$ to $1/10$ of the cell size, see figure~\ref{fig:applications}(b). Such patterns have been observed in L-form \cite{leav09,merc14} bacteria, as well as in coccal bacteria. \cite{zapu08} The patterns formed by PlsY and CdsA proteins (both essential to lipid metabolism) in \emph{Staphylococcus aureus}, in particular, show a striking coupling between protein density and membrane curvature. \cite{garc15} As seen in figures~\ref{fig:PvsK} and \ref{fig:PvsW}, and determined from (\ref{eq:Klines0}), modes with $\ell \leq 10$ are expected for $P>1.4 \cdot 10^{-4}$ which, for a typical cell size of $R=1~\mu$m, would correspond to correlation lengths $\xi > 10$~nm, well within the biologically plausible range.

\subsubsection{Nano-sized membrane rafts}

The existence of protein-rich \emph{raft domains} in the plasma membrane was controversial for some time, partly due to a conflation between the macroscale fluid-fluid phase separation observed in model lipid membranes with the observation of rafts in living cells. \cite{jaco07} Nevertheless, it is currently accepted that rafts are dynamic, fluctuating assemblies of proteins with sizes on the order of tens of nanometers. \cite{jaco07,ling10,simo10,sezg17} The precise physical mechanism behind raft formation, however, is still a matter of debate. Currently proposed theories include that rafts are compositional fluctuations near the critical point of fluid-fluid phase separation in lipid membranes, \cite{veat08} or that the actin cortex underlying the plasma membrane acts as a `picket-fence' which inhibits the lateral diffusion of proteins and promotes the formation of nano-scale aggregates. \cite{ritc03}

\begin{figure}[h]
\centering
\includegraphics[width=0.5\linewidth]{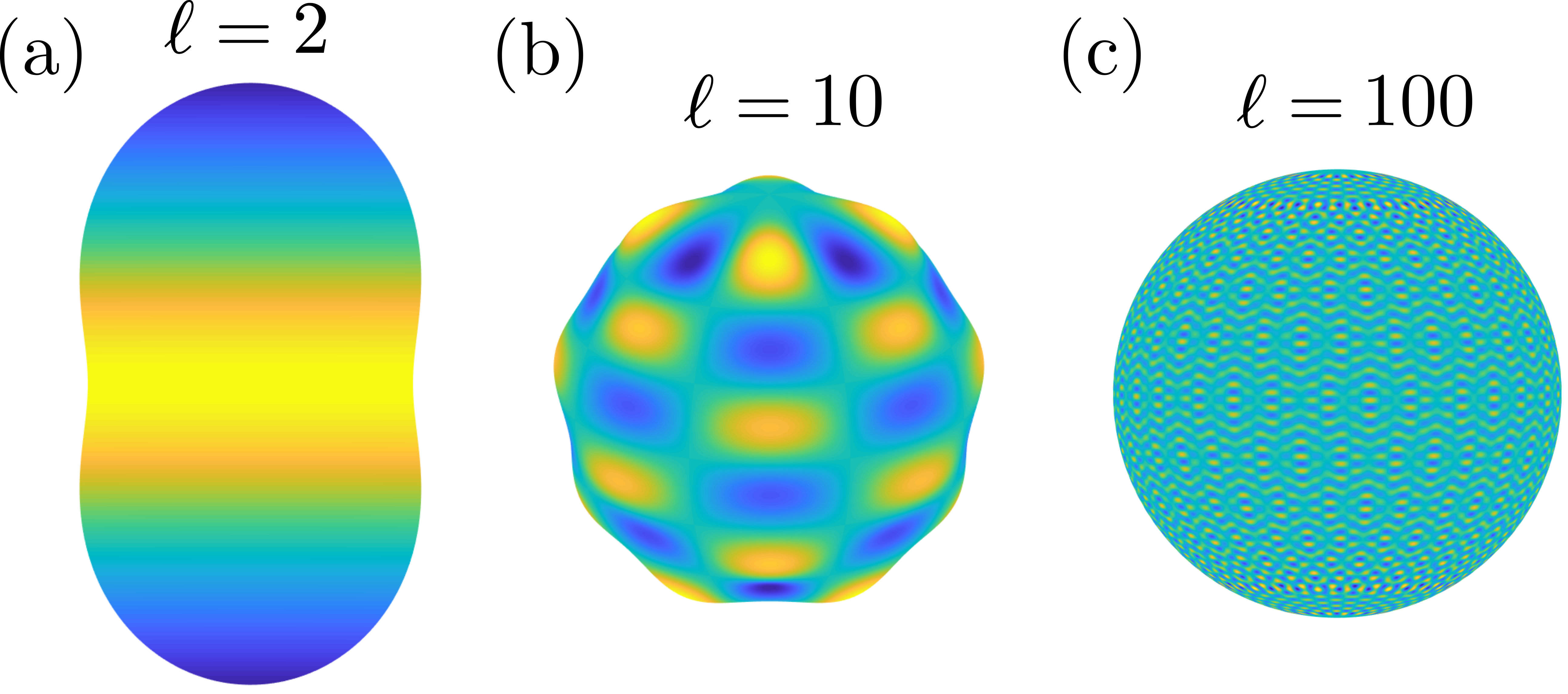}
\caption{Three examples of spontaneous pattern formation \emph{via} shape and protein distribution instability. (a) Instability $\ell=2$, corresponding to cell division with two distinct poles and an equatorial line. The mode $\ell=2$ can be triggered if $P>0.056$, see figures~\ref{fig:PvsK} and \ref{fig:PvsW}. (b) Instability $\ell=10$, corresponding to large-scale protein organisation such as that observed in \emph{Staphylococcus aureus}. \cite{garc15} The mode $\ell=10$ can be triggered if $P>1.4 \cdot 10^{-4}$. (c) Instability $\ell=100$, corresponding to the formation of nano-sized protein-rich membrane rafts. The mode $\ell=100$ can be triggered if $P>1.9 \cdot 10^{-8}$.  }
   \label{fig:applications}
\end{figure}

The model that we have presented here predicts that, under biologically reasonable parameters, curvature-inducing proteins can spontaneously self-organise into patterns that may be built from spherical harmonics with very high-order $\ell$-modes, with $\ell \gg 1$. As a consequence, in such cases the typical size of the protein-rich domains (which goes as $\sim R/\ell$) will be much smaller than the cell size $R$, leading to domain sizes on the order of tens of nanometers for a micron-sized cell, see figure~\ref{fig:applications}(c) for an example with $\ell=100$. It is therefore tempting to speculate that the mechanism presented here might also be connected to the existence of such nano-scale protein-rich rafts. Indeed, let us use the quantitative estimates of parameters obtained above, with typical values of membrane bending rigidity $\kappa=10~k_\mathrm{B}T$, correlation length of protein density fluctuations of $\xi=10$~nm, tethering strength of the membrane to the cell cortex $k_\mathrm{te} \approx 2.5 \cdot 10^6~k_\mathrm{B}T/\mu$m$^4$, and membrane tension $\sigma = 30$~pN/$\mu$m. For a small cell of radius $1~\mu$m, we can calculate our dimensionless parameters as $P=10^{-4}$, $K=2.5 \cdot 10^5$, and $T=7 \cdot 10^2$. Using these values in equation~(\ref{eq:Klines}), we expect the first unstable mode to be $\ell \approx 50$. This would correspond to a typical domain size $R/\ell \approx 20$~nm. For a larger cell of radius $10~\mu$m, we calculate $P=10^{-6}$, $K=2.5 \cdot 10^9$, and $T=7 \cdot 10^4$. Using these values in (\ref{eq:Klines}), we find $\ell \approx 500$ for the first unstable mode, once again corresponding to a typical domain size $R/\ell \approx 20$~nm.

\subsection{Summary}

To summarise, we have explored in detail pattern formation in spherical membranes that contain curvature-inducing proteins. Pattern formation arises from the interplay between membrane curvature energy, protein density fluctuations, and geometric constraints such as membrane tension and confinement forces due to the tethering of the membrane to the cell wall/cortex. We have shown that pattern formation in this system is controlled by just four dimensionless parameters, $W$, $K$, $T$, and $P$, defined in (\ref{eq:dimless}) and (\ref{eq:defW}). These parameters represent the number of curvature-induced proteins on the membrane, the confinement of the membrane due to the cell wall/cortex, the membrane tension, and the correlation length of protein density fluctuations, respectively. In most circumstances, pattern formation is expected to occur as the result of an increase in the average surface density of proteins (i.e.~the total number of proteins on the membrane surface), or of a relaxation of the geometric constraints on the membrane due to membrane tension or membrane tethering to the cell wall/cortex. The patterns that arise consist of protein-rich, highly-curved domains that alternate with protein-poor, weakly-curved domains. We hypothesise that spontaneous pattern formation as described here might be exploited by biological cells as a way to regulate their geometry in situations that require spatial organisation, symmetry breaking or polarisation of the cell, using only a minimal number of ingredients.

\section*{Acknowledgements}
We would like to acknowledge fruitful discussions with J.~Garcia-Lara and S.~Foster. This work was supported by the Human Frontiers Science Program (HFSP) RGP0061/2013. J.A-C.~acknowledges support from the Federal Ministry of Education and Research (BMBF, Germany) \emph{via} the consortium MaxSynBio, as well as from the Penn State MRSEC, Center for Nanoscale Science, under the award NSF DMR-1420620. Correspondence should be addressed to ramin.golestanian@physics.ox.ac.uk.

\bibliographystyle{ieeetr}
\bibliography{biblio}

\end{document}